\documentstyle[prl,floats,aps]{revtex}
\begin{document}
\draft
\twocolumn
\begin{title}
{Breakdown of the Migdal-Eliashberg theory in the strong-coupling
adiabatic regime}
\end{title} 
\author{A.S. Alexandrov}
\address
{Department of Physics, Loughborough University, Loughborough LE11 
3TU, United Kingdom}

\maketitle
\begin{abstract}
In view of some recent works on the role of  vertex corrections in the
electron-phonon system we readress an important
question of the validity of the Migdal-Eliashberg theory.
 Based on the solution of the Holstein model and
$1/\lambda$ strong-coupling expansion, we argue that  the
standard Feynman-Dyson perturbation theory by Migdal and Eliashberg
$with$ or without vertex corrections  
cannot be applied  if the electron-phonon  coupling is strong ($\lambda
\geq 1$) at any ratio of the phonon, $\omega$ and Fermi, $E_F$ energies. 
 In the extreme adiabatic limit ($\omega << E_F$) of the Holstein model electrons
 collapse into   self-trapped small polarons or bipolarons due
to spontaneous translational-symmetry breaking at
 $\lambda \simeq 1$. With the increasing phonon frequency the region
 of the applicability of the theory shrinks to lower values of
 $\lambda < 1$. 
 
\end{abstract}

\narrowtext

\noindent 
 The theory of ordinary metals is based on  'Migdal's'  theorem 
 \cite{mig}, which showed that the contribution of the diagrams with
 'crossing' phonon lines (so called 'vertex' corrections) is
 small. This is true if
 the parameter $\lambda \omega/ E_{F}$ is small. Neglecting the vertex corrections Migdal 
 calculated the renormalised electron mass  as $m^*=m(1+\lambda)$
 (near the Fermi level) \cite{mig}  and by breaking the gauge symmetry
 Eliashberg extended  the Migdal theory to describe the BCS superconducting state
 at  
 intermediate values of $\lambda$ \cite{eli}. The same theory, 
 applied to phonons, yields  the renormalised phonon frequency
 $\tilde{\omega}=
\omega (1-2\lambda)^{1/2}$ \cite{mig} with an instability at
 $\lambda=0.5$. Because of this instability both Migdal \cite{mig} and
 Eliashberg
\cite{eli} restricted  the applicability of their approach to $\lambda
 \leq 1$. 

It was then shown that if the adiabatic Born-Oppenheimer
approach is properly applied to a metal, there is only negligible renormalisation
of the phonon frequencies of the order of the adiabatic ratio,
$\omega/E_{F} <<1$ for $any$ value of $\lambda$  \cite{gei}. The
conclusion was that  the standard  electron-phonon
interaction could be applied to electrons for any value of $\lambda$ but it
should not be applied to renormalise phonons \cite{ref}. As a result, many
authors  used the Migdal-Eliashberg
theory with $\lambda$ much larger than 1 (see, for example, Ref.\cite{sca}). 

However,  starting from the
infinite coupling limit, $\lambda=\infty$ and applying  the inverse
($1/\lambda$) expansion   technique \cite{fir}  we showed \cite{ale,alemaz,ale0}  that the
many-electron system collapses into small polaron regime
at $\lambda \sim 1$ almost  independent of the adiabatic ratio. This regime is
beyond the Migdal-Eliashberg theory. It cannot be reached by summation of   
the standard Feynman-Dyson perturbation diagrams even including $all$
vertex corrections, because
of the broken
translational symmetry,  as first discussed by Landau \cite{lan}
for a single electron and by us \cite{alemaz} for the many-electron
system. 
During last years quite a few numerical and analytical  studies have
confirmed this conclusion
\cite{kab,kab2,bis,feh,mar,alemot,tak,dev,feh2,tak2,rom,lam,zey,wag,bon2,alekor,aub2,ale2}
(and references theirein).
On the other hand a few others (see, for example,
\cite{abr,dol}) still argue that the
breakdown of the Migdal-Eliashberg theory might  happen only  at $\lambda
\geq E_{F}/\omega >>1$. Indeed 
numerical study of the finite bandwidth effects \cite{alemaz2} and  some
analytical calculations of the vertex corrections to the vertex function \cite{all,pie,dol} with the standard Feynman-Dyson perturbation
technique  confirm the second conclusion.

In this letter I compare the Migdal solution of the Holstein
Hamiltonian with the exact one in the extreme adiabatic regime,
$\omega/E_{F} \rightarrow 0$, to show that the ground state of the
system is a self-trapped insulating state with broken translational
symmetry already
at
$\lambda \geq 1$.  

 The vertex corrections and the finite
 bandwidth are rather  technical issues, playing no role in the extreme
 adiabatic limit \cite{mig,all,alemaz2}. There is, however, another basic
 assumption of the 
 canonical Migdal-Eliashberg theory. That is   the electron and phonon Green
 functions (GF)  are translationally invariant. As a result one takes $G({\bf 
 r},{\bf r'},\tau)=G({\bf r}-{\bf r'},\tau)$ with  Fourier component
 $G({\bf k}, \Omega)$ prior to solving the Dyson equations.  This assumption excludes the 
 possibility of  local  violation of the  translational symmetry due to 
 the lattice deformation in any order of the Feynman-Dyson
 perturbation theory. This is  similar to  the absence
 of the anomalous (Bogoliubov) averages  in
 any order of the perturbation theory. To enable 
 electrons to relax into the lowest polaronic states, one has to introduce 
an infinitesimal translationally non-invariant 
 potential, which should be set equal to zero only in the final solution for the
 GF \cite{alemaz}. As in the case of the off-diagonal 
 superconducting order parameter, a small 
 translational-symmetry-breaking potential drives the system into a new ground state at 
 sufficiently large coupling, $\lambda \sim 1$, independent of the
 adiabatic ratio. Setting it equal to zero in the solution 
 of the equations of motion restores the translational symmetry but in a 
 new polaronic band rather than in the bare electron band, which turns out to 
 be an excited state. 

In particular,  let us consider  the
 extreme adiabatic limit of the Holstein chain \cite{hol}, which has
 the simple analytical solution:
\begin{eqnarray}
H &=& -t \sum _{<ij>} c_{i}^{\dagger}c_{j} +H.c. +2 (\lambda k
    t)^{1/2}\sum_{i} x_{i} c_{i}^{\dagger}c_{i}\cr
&+& \sum_{i}\left
    ( -{1\over{2M}} {\partial^2\over{\partial 
    x_{i}^2}}+{kx_{i}^2\over{2}} \right),
\end{eqnarray}
where $t$ is the nearest neighbour hopping integral, $c_{i}^{\dagger},
c_{i}$ are the electron operators, $x_{i}$ is the normal coordinate
of the molecule (site) $i$, and $k=M\omega^2$ with $M$ the ion mass \cite{ref2}.
We first consider  the two-site case
(zero dimensional limit), $i,j=1,2$ with one electron, and than generalise the result for the
infinite lattice with many electrons.  The transformation $X=(x_1
+x_2)$, $\xi=x_1 -x_2$ allows us to eliminate the coordinate $X$,
which is coupled only  with the total density ($n_1+n_2=1$), leaving the
following Hamiltonian to solve in the extreme adiabatic limit $M
\rightarrow \infty $:
\begin{equation}
H= -t(c_1^{\dagger}c_2+c_2^{\dagger}c_1) + (\lambda k
    t)^{1/2} \xi( c_{1}^{\dagger}c_{1}-
    c_{2}^{\dagger}c_{2}) +{k\xi^2\over{4}}.
\end{equation}
The solution is  
\begin{equation}
\psi= (\alpha c_1^{\dagger}+\beta c_{2}^{\dagger}) |0>,
\end{equation}
where
\begin{equation}
\alpha= {t\over{[t^2+((\lambda k t)^{1/2}\xi+(t^2+\lambda kt \xi^2)^{1/2})^2]^{1/2}}},
\end{equation}
\begin{equation}
\beta=- {(\lambda k t)^{1/2}\xi+(t^2+\lambda kt \xi^2)^{1/2}\over{[t^2+((\lambda k t)^{1/2}\xi+(t^2+\lambda kt \xi^2)^{1/2})^2]^{1/2}}},
\end{equation}
and the energy 
\begin{equation}
E= {k\xi^2\over{4}}- (t^2+\lambda k t \xi^2)^{1/2}.
\end{equation}

In the extreme adiabatic limit the displacement $\xi$ is classical, so
the ground state energy, $E_0$ and the ground state displacement $\xi_0$ are obtained by minimising
Eq.(6) with respect to $\xi$. If $\lambda \geq 0.5$ one obtains 
\begin{equation}
E_0=-t(\lambda+ {1\over{4\lambda}}),
\end{equation}
and
\begin{equation}
\xi_0=\left[{t(4\lambda^2-1)\over{\lambda k}}\right]^{1/2}.
\end{equation}   
  The   symmetry-breaking ('order') parameter is
\begin{equation}
\Delta\equiv \beta^2-\alpha^2= {[2\lambda +(4\lambda^2-1)^{1/2}]^{2}-1\over{[2\lambda +(4\lambda^2-1)^{1/2}]^{2}+1}}.
\end{equation}

If, however, $\lambda<0.5$ 
the ground state is translationally invariant with
$E_0=-t, \xi=0, \beta=-\alpha$, and $\Delta=0$. Precisely this state
is the 'Migdal' solution of the Holstein model. Indeed, in the
Migdal approximation GF is diagonal in the ${\bf k}$ representation,
$G({\bf k,k'},\tau)= G({\bf k}, \tau) \delta_{\bf
k,k'}$. The site operators can be transformed  into   momentum space as 
\begin{equation}
c_{k}= N^{-1/2}\sum_{j} c_{j} \exp(ikaj),
\end{equation} 
with $a$ the lattice constant, $N$ the number of sites, and $k=2\pi
n/Na$ with $ -N/2< n \leq N/2$.  Than the off-diagonal GF with
$k=0$ and $k'=\pi/a$ of the two-site chain ($N=2$) at $\tau=-0$ is
given by
 \begin{equation}
G(k,k',-0)={i\over{2}}\langle
(c_1^{\dagger}-c_2^{\dagger})(c_1+c_2)\rangle.
\end{equation}
Calculating this average one obtains
\begin{equation}
G(k,k',-0)={i\over{2}}(\alpha^2-\beta^2),
\end{equation}
which should vanish in the Migdal theory. Hence, this theory only
provides symmetric  (translationally invariant) solution with
$|\alpha|=|\beta|$. When $\lambda>0.5$ this solution is $not$ the
ground state of the system, Fig.1. The system collapses into a
localised adiabatic polaron, trapped on the 'right' (or on the 'left')
site due to a finite local lattice deformation $\xi_0$.  On the other hand, when
$\lambda<0.5$, the Migdal solution is the $only$ solution  with $\xi_0=0$. Thus  the Migdal-Eliashberg constraint \cite{mig,eli} on the applicability of
their approach is perfectly correct  irrespective of the phonon frequency
renormalisation. 

The generalisation to the multi-polaron system on the infinite
lattice of any dimension is straightforward in the extreme adiabatic regime. The adiabatic
solution of the   infinite one-dimensional (1D) chain with one
electron was
obtained by Rashba \cite{ras} in the continuous approximation, and by 
Holstein \cite{hol} and Kabanov and Mashtakov \cite{kab} for a discrete
lattice. The last authors also studied  the Holstein two-dimensional (2D) and
three-dimensional (3D) lattices in the adiabatic limit.
According to Ref. \cite{kab} the self-trapping  of a single
electron occures for any value of $\lambda$ in a 1D Holstein chain, and 
at $\lambda \geq 0.875$ and $\lambda \geq 0.92$ in 2D and 3D,
respectively. The single-polaron GF is not translationally
invariant in this strong-coupling adiabatic limit. The radius of the self-trapped adiabatic polaron, $r_{p}$, is
readily derived from its continous wave function \cite{ras}
\begin{equation}
\psi(x) \sim 1/\cosh(\lambda x/a).
\end{equation}
It becomes smaller than the lattice constant, $r_{p}=a/\lambda$ for
$\lambda\geq 1$. That is why in the strong-coupling ($\lambda \geq 1$)
adiabatic regime  
the multi-polaron  system remains in the self-trapped
insulating state no matter how many polarons it has. The only
instability which might occur in this regime is the formation of
on-site self-trapped bipolarons, if the Holstein
 on-site attractive 
interaction, $2 \lambda zt$, is larger than the repulsive Hubbard $U$
\cite{and}.  Actually, this instability can be seen  in the second
order
Feynamn-Dyson diagramm containing the polarisation loop, as explained
in Ref. \cite{kab3}. On-site   bipolarons forme
charge ordered insulating state due to  weak repulsion between them \cite{aleran}. The
exact analytical and numerical  proof of this statement as well as  different
 polaronic and bipolaronic configurations in the adiabatic Holstein model  was reviewed by
Aubry \cite{aub}. For example, the asymptotically exact many-particle
ground state of the half-filled Holstein model in the
strong-coupling limit ($\lambda >> 1$) is 
\begin{equation}
\psi= \prod_{j\in B}c^{\dagger}_{j,\uparrow}c^{\dagger}_{j,\downarrow}
|0 \rangle
\end{equation}
for any value of the adiabatic ratio, $\omega/zt$
\cite{aleran,aub}. Here ${j}$ are $B$-sites of the bipartite lattice $A+B$.  It is a charged ordered insulating state,
rather than the Fermi liquid, expected in the Migdal approximation at
any value of $\lambda$ in the extreme adiabatic limit, $\omega
\rightarrow 0$. If  the Coulomb repulsion (Hubbard $U$)
is sufficiently 
strong to prevent the bipolaron formation,   than at half filling every
site is occupied by one polaron. The deformation barrier for their
tunneling to a neighboring site disappears, so one could erroneously
believe that the system should be metallic. However, to keep polarons
unbound into on-site bipolarons, the Hubbard $U$ should be larger than
$2\lambda zt$, i.e. larger than the bandwidth, $U> 2zt$, if $\lambda
>1$. It is well known that in this regime  the system is a
Mott-Hubbard insulator, rather than an ordinary metal. Hence,  the Migdal-Eliashberg
theory cannot be applied at $\lambda \geq 1$, no matter what a
strength of the Coulomb
interaction, the  number of electrons and the dimension of the lattice
are.    

The non-adiabtic corrections (phonons) allow 
polarons and bipolarons propagate as  Bloch
waves in a new (narrow) band.  Thus, under   certain conditions \cite{ale2} the multi-polaron system
might be metallic with  polaronic (or bipolaronic)
carriers rather than  bare electrons. However,  there is a
qualitative difference between the ordinary Fermi liquid   and the
polaronic one. In particular,  the renormalized
(effective) mass of electrons is
independent of the ion mass $M$ in ordinary metals (where the Migdal
adiabatic approximation is believed to be valid), because $\lambda$
does not depend on the isotope mass.   However, the polaron  effective mass $m^{*}$ will depend on
$M$. This is because the polaron mass $m^{*}= m \exp (A/\omega)$
\cite{ale3}, where $m$ is the band mass in the absence of the electron-phonon
interaction, and  $A$ is a constant.  Hence, there
is a large isotope effect on the carrier mass in polaronic metals, in
contrast to the zero isotope effect in ordinary metals. Recently,  this effect has
been experimentally found in cuprates \cite{mul} and manganites
\cite{pet}. With the increasing phonon frequency  the $1/\lambda$ polaron
expansion
becomes valid for a smaller values of $\lambda$ (see, for example,
\cite{ale2}). Hence, the region of the applicability of the Migdal
approach shrinks with increasing $\omega$ to the smaller values of the
coupling, $\lambda<1$. 

The essential physics of   strongly coupled
electrons and phonons  has been understood with the $1/\lambda$
expansion technique \cite{ale0,alemot}, which starts with  the exact
solution in the extreme
limit $\lambda=\infty$ and  allows for the summation of all multiphonon
diagrams in any order of the small parameter $1/\lambda$ for any value
of the adiabatic ratio $\omega/t$. It predicts a breakdown of
the Migdal-Eliashberg theory at $\lambda \simeq 1$ or less, in agreement with the exact solution of the extreme
adiabatic Holstein model discussed here. There are also other
studies of the same problem, which do not relay on the standard
Feynman-Dyson perturbation theory in powers of $\lambda$. In
particular, Takada $et$ $al$  \cite{tak,tak2,tak3} applied the
gauge-invariant self-consistent method 
neglecting the vector term in the Ward identity. Benedetti and
Zeyher \cite{zey} applied the dynamical mean-field theory in infinite
dimensions. As in  the $1/\lambda$ expansion technique,  both approaches
avoided the problem of the broken translational symmetry by
using the  nondispersive vertex and Green functions as the starting point. As a result
they  arrived at the same correct conclusion about  the applicability of the
Migdal approach (in Ref. \cite{zey} the critical value of $\lambda$
was found to be 1.3 in the adiabatic limit). Meanwhile, some
authors \cite{dol,abr}, who  relay on the perturbation expansion in
powers of $\lambda$, fail to
notice the self-trapping transition, erroneously claiming
that the
Migdal-Eliashberg theory could be applied at practically any $\lambda$
in the extreme  adiabatic regime.  

In conclusion, I have shown that the Migdal-Eliashberg theory cannot
be applied  in the   adiabatic regime of the
strongly coupled many-electron system if the (BCS) coupling
constant, $\lambda$,  is about  1 or larger irrespective of the number
of electrons. There is a
transition into the self-trapped state due to the broken translational
symmetry, Fig.1. The transition appears at $0.5 <\lambda < 1.3$
depending on the lattice dimensionality. Naturally, with the increasing
phonon frequency the region of the applicability of the Migdal
approach shrinks further to lower values of $\lambda$. I believe that this
conclusion is correct for any electron-phonon interaction conserving
the on-site electron  occupation numbers. In particular, 
  Hiramoto and Toyozawa \cite{toy} calculated the 
  strength of the deformation potential, which transforms electrons
  into  small polarons
  and bipolarons. 
 Their continuous approach 
  is sufficient for a  qualitative estimate if 
  the Debye  wavenumber $q_{D}\sim \pi/a$ is introduced as an upper 
  limit cut-off in all sums in the
  momentum space.  Hiramoto and 
  Toyozawa found that
   the transition between two  electrons and a small bipolaron 
   occurs at $\lambda\simeq 0.5$, that is half of the 
  critical value of $\lambda$ at which the transition from 
  electron  to small polaron takes place
 in the extreme adiabatic limit 
  ($sq_{D}<<zt$, s the sound velocity). The effect of the adiabatic ratio $sq_{D}/zt$ on the 
  critical value of $\lambda$ was found to be negligible. The radius
  of the acoustic polaron and bipolaron is  the lattice
  constant, so the critical $\lambda$ does not depend on the number
  of electrons in this case either.  
  
The author greately appreciates enlightening discussions with A.R. Bishop,
V.V. Kabanov, R. Zeyher, and J.R. Schrieffer.

{\bf Figure caption}

\vspace{0.5cm}

Fig.1. The ground state energy (solid line), $E/t$, and the
'order' parameter (thin line), $\Delta$, of the adiabatic Holstein model as functions
of the coupling constant, $\lambda$. There is the symmetry breaking transition at
$\lambda=0.5$. The energy of the symmetric state is shown by the  dashed line.

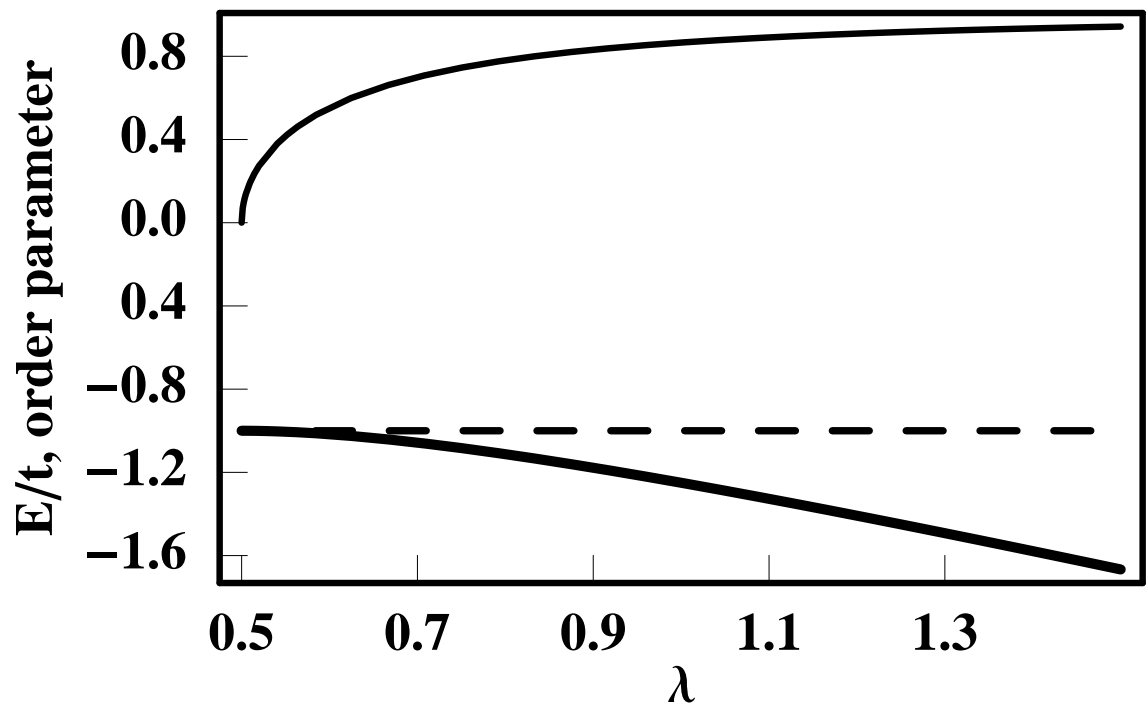
\end{document}